# Quantum assisted trustworthiness for the Quantum Internet

Agustín Zaballos, Adrià Mallorquí and Joan Navarro

*Abstract*—Device redundancy is one of the most well-known mechanisms in distributed systems to increase the overall system fault tolerance and, consequently, trustworthiness. Existing algorithms in this regard aim to exchange a significant number of messages among nodes to identify and agree which communication links or nodes are faulty. This approach greatly degrades the performance of those wireless communication networks exposed to limited available bandwidth and/or energy consumption due to messages flooding. Lately, quantum-assisted mechanisms have been envisaged as an appealing alternative to improve the performance in this kind of communication networks and have been shown to obtain levels of performance close to the ones achieved in ideal conditions. The purpose of this paper is to further explore this approach by using super-additivity and superposed quantum trajectories in quantum Internet to obtain a higher system trustworthiness. More specifically, the wireless communication network that supports the permafrost telemetry service for the Antarctica together with five operational modes (three of them using classical techniques and two of them using quantum-assisted mechanisms) have been simulated. Obtained results show that the new quantum-assisted mechanisms can increase the system performance by up to a 28%.

## I. INTRODUCTION

DESPITE extensive efforts to expand mobile network coverage worldwide through wireless communication technologies (such as 3G, 4G, LTE, and 5G), there are still regions and specific use cases that remain unable to harness the benefits of these technologies due to their intrinsic characteristics. For example, when providing reliable connectivity in remote areas poses challenges as conventional networks often fail to reach these locations, thereby necessitating more expensive alternatives like satellite communication for users in such regions. Similarly, Internet of Things (IoT) devices exposed to diverse and sometimes harsh environmental conditions and necessitated of low-power and long-range communication are typically forced to rely on well-known technologies like Narrowband IoT (NB-IoT) or Long-Range Wide Area (LoRa).

To address any potential limitations of these communication technologies and explore alternative solutions, the SHETLAND_NET research project [1] seeks to establish an experimental wireless communication network in Antarctica to support various in-field experiments. This network comprises two main layers: an access layer utilizing LoRa to connect different sensing devices, and a backhaul employing Near Vertical Incidence Skywave (NVIS) technology to provide coverage for various IoT domains within the South Shetland Islands archipelago [2]. Currently, the scientific experiments conducted on top of this communication network encounter challenges such as critical data losses, experiment delays, and costly troubleshooting efforts. These issues primarily stem from two factors: 1) the inherent unreliability of NVIS, which is highly contingent on ionosphere conditions and solar activity, and 2) the inability of the deployed devices to endure the extreme environmental conditions of Antarctica resiliently. Consequently, the overall system becomes unreliable and untrustworthy, often rendering the collected data unusable for scientific purposes.

Recently, the usage of quantum mechanics to improve the performance of this communication network has been investigated [3]. More specifically, conducted simulations have shown that quantum-assisted mechanisms could improve the overall system trustworthiness by taking advantage of devices redundancy and obtaining a performance—measured in terms of Successful Transaction Rate (STR)— that closely approaches the performance achieved by ideal centralized reputation techniques facilitated by a social IoT trustworthiness layer [3][4].

This paper proposes a new quantum-assisted mechanism based on super-additivity and superposed trajectories for quantum Internet to further increase the system's trustworthiness and, thus, the STR. To contextualize the simulations, an IoT permafrost telemetry service deployed on top of the aforementioned Antarctic wireless communications network has been considered. According to our experience in the SHETLAND-NET research project, the target STR for a successful operation of the permafrost telemetry must be at least 60% [5]. In the proposed simulation scenario of this work, the STR is also used to evaluate the goodness of the proposed quantum-assisted mechanisms and to anticipate and identify possible weaknesses in an IoT telemetry system. This research offers, by means of simulations, the opportunity to explore potential advantages of the proposed quantum-assisted mechanisms regarding social IoT trustworthiness layer and to evaluate the benefits of altering measuring spots redundancy over the quantum Internet [6].

Agustín Zaballos, Adrià Mallorquí and Joan Navarro are with the Engineering Department of La Salle Campus Barcelona (Universitat Ramon Llull – URL), Barcelona, Spain. E-mail: agustin.zaballos@salle.url.edu, adria.mallorqui@salle.url.edu jnavarro@salle.url.edu.

## II. TRUSTWORTHINESS THROUGH REDUNDANCY

Trustworthiness in a communication network refers to the degree of reliability of the network's nodes and processes in exchanging information. Redundancy, in the context of trustworthiness, is related to the presence of multiple redundant measuring spots or communication paths within the data network. Redundancies serve as a backup that ensures that if one measuring spot or path becomes compromised, other reliable options are available to maintain uninterrupted communication.

The trustworthiness of the Internet is essential for proper functioning under harsh conditions. The literature defines Internet trustworthiness through four dimensions [3][4][7][8]. Hence, the baseline trustworthiness model proposed to evaluate the Antarctic quantum Internet's performance is layer-based and comprises four layers: data trustworthiness, network trustworthiness, social trustworthiness, and consensus layers. The data trustworthiness layer ensures the accuracy of the data provided by the source, the network trustworthiness layer ensures that packets reach their destination unaltered and on time, the social trustworthiness layer leverages objects' social relationships to improve trust, and the consensus layer ensures that all participants agree on the same data values through a general agreement. The permafrost-related experiment in which this research is contextualized is focused on the goodness assessment of measuring spots redundancy. Measuring spots redundancy is necessary for the last two layers if we want to maximize their usefulness.

The social trustworthiness layer essentially focuses on establishing trust among network nodes through reputation-based mechanisms. These mechanisms evaluate the trustworthiness of nodes based on factors such as previous transaction feedback, indirect opinions from other nodes, transaction relevance, node centrality, computational capacity, and the nature of relationships between nodes [8]. It aims to determine which nodes are trustworthy for information exchange, maximizing the number of successful transactions. On the other hand, the consensus layer aims to achieve a decentralized general agreement among all participating nodes. It employs a voting-based mechanism that tolerates a certain percentage of byzantine nodes. Understanding the dependencies between these two layers is essential for evaluating and quantifying the trustworthiness of the communication network.

The proposed quantum trustworthiness model combines reputational and consensus quantum algorithms for optimal trustworthiness performance. [3] explores the use of the Fast Quantum Consensus algorithm to achieve instantaneous coordination of communication parties, avoiding traffic congestion's negative impact. In traffic-overwhelmed scenarios, voting-based consensus algorithms can lead to increase traffic congestion, decreasing the Internet's trustworthiness. [3] concludes that implementing a quantum consensus management plane improves system trustworthiness by taking advantage of measuring-spots redundancy with an improvement of 16% compared to its classical alternative. It turns out that distributed fast quantum consensus barely improves by 5% the reference value obtained through centralized reputation algorithms (social trustworthiness layer over classical Internet).

## III. QUANTUM-ASSISTED TRUSTWORTHINESS

Given the observed improvement in trustworthiness achieved by quantum consensus compared to classical consensus (specifically, Fast Quantum Consensus over the Practical Byzantine Fault Tolerance protocol), and with the aim of getting closer to the maximum reference value provided by the classical social trustworthiness layer [4], this paper poses the following research question: What enhancements could be achieved by applying quantum-assisted mechanisms, such as super-additivity and superposed trajectories, to the social trustworthiness layer, which sets an upper bound on STR in the classical Internet?

In this way, our objective is to explore the potential benefits derived from incorporating quantum features into the social trustworthiness layer. To this end, we have evaluated how leveraging quantum properties, such as super-additivity and superposed trajectories, can further improve the performance and reliability of communication protocols beyond what is currently attainable in classical systems. By doing so, we aim to provide valuable insights into the feasibility and potential advantages of integrating quantum enhancements, thereby paving the way for advancements in Antarctica telemetry and potentially other domains where reliability and efficiency are critical.

Getting into the matter, quantum super-additivity refers to the phenomenon where the total amount of information that can be transmitted through a quantum channel is greater than the sum of the individual capacities of its subchannels. In other words, by combining two or more quantum channels, one can achieve a greater amount of information transmission than by using the channels separately [9]. This concept is related to the properties of entanglement in quantum mechanics. Entanglement is a type of correlation between quantum systems that can exist even when they are physically separated. This means that manipulating the entangled systems in a certain way makes it possible to transmit more information than can be achieved by transmitting the systems separately. In fact, when two entangled qubits are sent through separate channel events, the amount of information that can be transmitted through each channel utilization individually is limited by the channel's capacity. However, by combining both channel events, the entanglement between the qubits can increase the total amount of information that can be transmitted. The phenomenon of quantum super-additivity has been demonstrated experimentally and has important implications for quantum communication and information theory [9].

On the other hand, quantum trajectories refer to a theoretical framework for describing the time evolution of qubits in terms of individual trajectories in the space of quantum states. In traditional quantum mechanics, the evolution of a qubit is defined by a wave function, which gives the probability amplitude for the system to be in a particular state. Quantum trajectory theory extends the wave function description by introducing the concept of quantum state reductions, which occur when a qubit interacts with its environment. In [10], authors demonstrate experimentally that superposing multiple quantum trajectories can enhance the performance of quantum communication protocols. They show that combining multiple trajectories in a qubit can increase the amount of information that can be transmitted through a quantum channel. The key is that by superposing multiple quantum trajectories, it is possible to exploit the inherent quantum mechanical properties of entanglement and nonlocality to enhance the performance of quantum communication protocols.

Both quantum mechanisms are a consequence of the non-classical correlations that can exist between different quantum channels' events. The success probability of a quantum channel's event could be enhanced by the existence of another entangled quantum channel's event. This means, for example, that by adding a weak channel to a strong channel, the capacity of the strong channel can be increased. In fact, this can be used in quantum channels in serial by transmitting a quantum state one after another. This can increase the capacity of the quantum channel and reduce the impact of noise or errors introduced by the environment. It highlights the non-classical properties of quantum channels and the potential for quantum protocols to achieve higher trustworthiness than classical protocols.

To implement super-additivity and superposition of trajectories in serial or parallel quantum communications, it is necessary to design protocols that can exploit the non-classical properties of quantum channels, such as entanglement and non-additivity [11]. These protocols can be complex and may require advanced quantum technology, but they have the potential to improve the performance of quantum communication in serial significantly. Although the fundamental quantum Internet technology is not yet mature, research on quantum Internet could be conducted. It should be said that while the physical and link layers of quantum networks require experimental assessments on costly hardware, upper-layer protocols can be tested via simulations when the simulation environment is consistently modeled with the practical physical facts [12].

This suggest that by leveraging these unique properties of quantum systems, it should be possible to develop a new and more efficient social trustworthiness layer for secure communication.

## IV. SIMULATION SETUP

The Riverbed Modeler simulator [6] has been used to model and evaluate our permafrost telemetry scenario. To conduct simulation tests and assess the outcomes using our proposed quantum-assisted trustworthiness model, we have modeled communication elements, including the classical communication media, such as the physical and link layers of LoRa and NVIS technologies. Moreover, we have incorporated the link layer mechanisms of the quantum Internet, including quantum pre-processing, quantum post-processing, and entanglement generation and distribution. The modeled scenario also includes the telemetry application layer, the detection of faulty behavior of byzantine nodes, and the management of social trust. All of them integrated with the quantum management plane and the quantum trustworthiness management plane (see Fig. 1). Because of the variability of NVIS availability, which is highly correlated with the ionosphere state and solar activity, a delay tolerant network has been implemented. This ensures that the network can operate efficiently within the range of 70% to 100% NVIS availability. In [8], finite state diagrams and the code for modeling our Antarctic experiment are described.

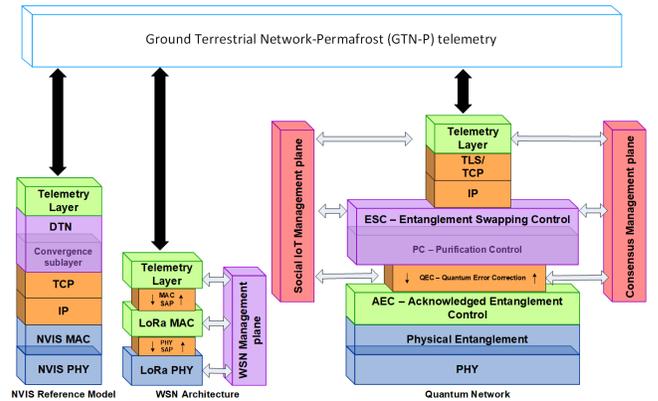

**Fig. 1.** Protocol architecture proposed for our experimentation.

In this regard, a reputation mechanism has been employed to model the social trustworthiness layer, aiming to identify the reliable nodes involved in redundant telemetry. Within this layer, concentrator elements calculate reputation by considering the feedback received from previous transactions, forming both short-term and long-term opinions [7]. In our simulations, all transactions are assumed to have equal weight, and the sensors are initialized in a similar manner. By utilizing a centralized and bandwidth-efficient implementation within the simulation scenario, we can infer the maximum achievable trustworthiness from a classical Antarctic network deployment with redundancy. It is important to note that the performance indicator used by this layer to assess its contribution to trustworthiness is also the Successful Transactions Rate (STR).

The consensus trustworthiness layer has been implemented to establish a shared state among all participants involved in redundant telemetry, pursuing a general agreement on a particular measurement. The implemented consensus protocol, widely used in IoT environments [4][8], uses a voting-based mechanism to tolerate byzantine nodes. While this mechanism contributes to enhancing trustworthiness in deployments with measuring-spots redundancy, it is essential to consider potential bandwidth consumption issues caused by the voting process. In situations of excessive traffic jams, particularly with a high number of redundant sensors, the system may experience congestion. To alleviate this effect in our simulated scenario, we have employed the Practical Byzantine Fault Tolerance mechanism [13].

The Quantum Consensus Management plane of Fig. 1, which is responsible for developing a quantum protocol to achieve a decentralized general agreement while minimizing message exchanges within the voting-based mechanism, is implemented through the Fast Quantum Consensus approach, as described in the experiment published in [14][15]. On the other hand, the Quantum Social Management plane is in charge of the sensor reputation management by using the super-additivity property and superposing multiple quantum trajectories as referenced experiments [9][10].

Finally, the physical and link layers have been modeled to be aligned with quantum mechanics [11] by enabling the generation and distribution of entangled pairs. The routing protocol incorporates a logical point-to-point quantum topology (i.e., path optimization based on quantum metrics is unnecessary). Considering its similarities to circuit-switched schemes, the transport layer incorporates a proactive congestion control mechanism for the quantum Internet. In short, the parameters employed for the classical Internet, specifically corresponding to LoRa and NVIS deployment, are promoted from the ones outlined in [4][8]. In the simulated scenarios, the main goal is to assess the impact of quantum-assisted mechanisms to approach the optimal global figure of merit (STR). Thus far, the classical social layer has proven to be the paradigm yielding the highest trustworthiness in the Antarctic telemetry use case.

The simulation scenario consists of five NVIS concentrator nodes, each including its own LoRa coverage area and equipped with sensors for telemetry. To evaluate the effectiveness of the quantum social and quantum consensus layers, redundancy in the measuring spots has been simulated at each GTN-P node to capture multiple measurements of the same data. Furthermore, we have introduced up to five additional redundant sensor nodes at each measuring spot, allowing for the tolerance of a single byzantine node. Our Antarctica use case experiment involves a simulation scenario with 32 and 64 permafrost measuring spots, enabling us to assess the capabilities of quantum-assisted mechanisms. Each test has been repeated 30 times, resulting in a total of up to 115,000 tests in which the average value of the STR has been computed. The obtained results that are exhibited have a confidence interval of 99%.

## V. RESULTS AND CONCLUSIONS

After the completion of the simulations, the STR was calculated by averaging the obtained values. The STR was assessed over a simulation duration of 400 days, which represents the period between two consecutive Antarctic campaigns that take place once per year during the southern summer. In addition, the simulated scenarios include various values for the byzantine fault probability ($Pb_0$), ranging from $10^{-1}$ to $10^{-3}$. These values allowed us to simulate the impact of utilizing different sensor sources and battery charge levels on the system's trustworthiness.

There are five different operational modes based on the utilization of redundant-related mechanisms. These five modes, as well as their combination, represent different strategies for handling redundancy and ensuring trustworthiness in the system: 1) In the "Standard" mode, nodes transmit data directly to the control center immediately after gathering it, without any further interaction. It is important to note that this approach does not involve using the social or consensus mechanisms and therefore does not capitalize on the redundancy of measuring spots. 2) In the "Social" mode, nodes incorporate the social trustworthiness layer. This means that nodes with a low reputation, as they often provide inaccurate data, are excluded from participating in the telemetry service by filtering out untrustworthy nodes. 3) In the "Consensus" mode, nodes reach a general agreement on the correct sensed value before transmitting it to the control center, despite the presence of potential byzantine nodes (i.e., thanks to Practical Byzantine Fault Tolerance protocol [4][8]. 4) In the "Quantum Consensus" mode, nodes also pursue a general agreement on the correct value, but they employ the Fast Quantum Consensus protocol [14][15]. This quantum-assisted approach leverages the advantages of quantum mechanics to minimize the number of communication packets required for achieving consensus. 5) Finally, in the "Quantum Social" mode, nodes develop the social trustworthiness layer by incorporating super-additivity and superposed multiple quantum trajectories to get closer to the theoretical optimum value of the total figure of merit, as explained before [9][10].

Figures 2 to 4 illustrate the overall pattern of STR values observed in our simulation results. In each figure, a grid with an X × Y dimension is created, representing all possible combinations of simulation parameters. Here, X represents the number of different $Pb_0$ values (i.e., byzantine fault probability), and Y represents the number of different sensors (i.e., measuring spots and redundancy). For each point on this grid and each mode combination, the average value of the trustworthiness STR metric is calculated. By connecting the STR values for adjacent points on the grid, a mesh is formed that encompasses all the STR values. This trustworthiness mesh provides a comprehensive representation of the trustworthiness across different simulated scenarios.

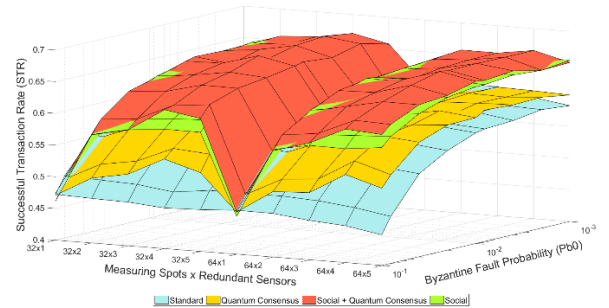

**Fig. 2.** Colored mesh for STR analysis (classical Internet and quantum-assisted consensus).

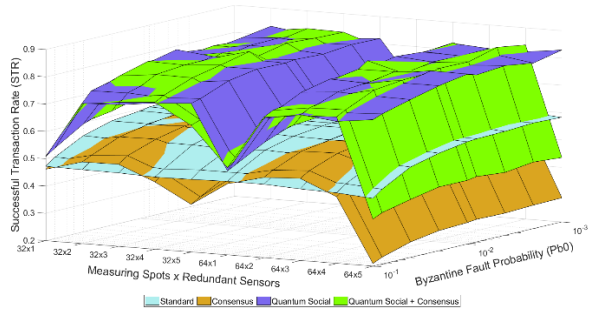

**Fig. 3.** Colored mesh for STR analysis (classical Internet and quantum-assisted social trustworthiness).

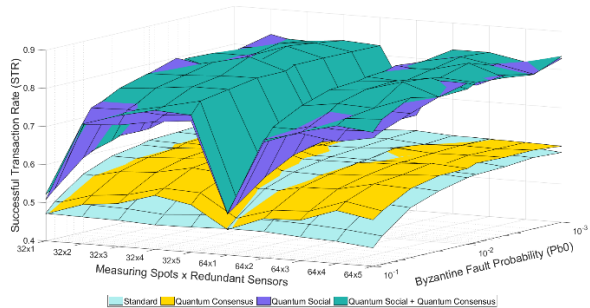

**Fig. 4.** Colored mesh for STR analysis (classical baseline and quantum-assisted mechanisms combination).

In general, as the "Byzantine Fault Probability" increases along its axis, the STR decreases because a higher $Pb_0$ leads to more faulty sensed values. On the other hand, as the "Redundant Sensors × Sensor Clusters" increases along its axis, the STR also decreases because of the introduction of more devices to the network, resulting in more packet losses due to network congestion. Obtained results demonstrate the advantages of our proposal within the practical limitations of our use case in terms of the number of measuring spots and sensor redundancy for the permafrost telemetry use case (i.e., the number of redundant sensors per cluster is limited to 5, so only one byzantine node is theoretically tolerated per consensus group and the simulation scenario includes 32 and 64 permafrost measuring spots). In this way, the "Redundant Sensors × Sensor Clusters" axis has 10 discrete points labeled as [32 × N, 64 × N], where N ranges from 1 to 5. The observed behavior remains consistent, as the STR values recover when transitioning from the "32 × 5" to the "64 × 1" point due to fewer nodes being introduced, resulting in fewer packet losses caused by network congestion. Once the method for interpreting Figures 2 to 4 is clear, one can easily understand the qualitative results of the tests. Figures 2 to 4 depict the trustworthiness mesh for the analyzed modes.

These results can be applied to other use cases as well. In fact, the obtained results demonstrate the advantages of our proposal within the practical limitations of our specific use case, which involves a limited number of measuring spots and sensor redundancy for permafrost telemetry. Therefore, the benefits of our proposal could be amplified if we extend the application to a general-purpose telemetry case over the

Internet in scenarios where a larger number of sensors can be deployed, and the required STR justifies greater redundancy. For instance, in the studied permafrost telemetry use case, the number of redundant sensors is restricted to 5, allowing for the tolerance of only one byzantine node per consensus cluster. However, we can further evaluate the improvement in trustworthiness offered by quantum consensus by increasing the number of redundant sensors. Figure 5 presents the simulation results when the number of redundant sensors is extended from 4 to 10, increasing the number of tolerated byzantine nodes from 1 to 3. This figure compares the results between the quantum-assisted simulated modes and the classical ones (i.e., we can observe the impact of quantum-assisted modes in contrast to classical modes).

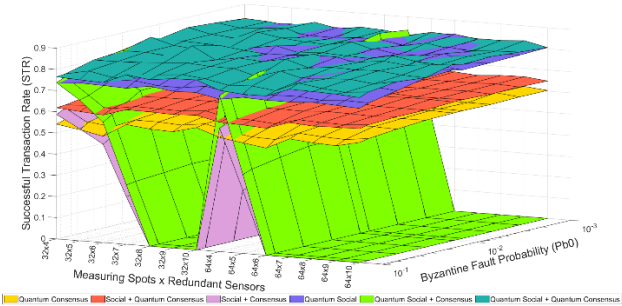

**Fig. 5.** Trustworthiness mesh varying the number of redundant sensors from 4 to 10.

In this generalization, it becomes evident that utilizing 5 or more redundant sensors with classical consensus leads to a degradation in the achieved STR, dropping below 0.6. Furthermore, the STR reaches 0 when there are more than 7 redundant sensors and 32 measuring spots, or more than 5 sensors and 64 measuring spots. This fact happens due to the exponential increase in complexity and the number of messages exchanged among group members to reach a general agreement, resulting in congestion within the low-bandwidth network. In contrast, the quantum consensus mechanism avoids this exponential growth in complexity by minimizing the number of exchanged messages [3]. As a result, the network can remain uncongested even when more nodes are utilized, thereby maintaining the obtained STR at values above 0.6. This quantum approach also slightly improves the overall trustworthiness compared to the reference mechanism based on the centralized social trustworthiness layer. Even more impressive is the result obtained using the quantum social mode. While we were previously concerned about losing an acceptable STR level for the permafrost use case (i.e., 60%) due to channel capacity, this new mode exceeds expectations by far. Although the quantum consensus reaches STR levels close to the reference value (i.e., the achieved with the social trustworthiness mode), the quantum social mode goes beyond the maximum reference value by over 27%, thus raising the reference value from a 60% STR to 85% in the simulated scenario.

|  | Maximum (use case) | Average (use case) | Maximum (extended) | Average (extended) |
|---|---|---|---|---|
| Standard | 0,610 | 0,562 | 0,610 | 0,550 |
| Social | 0,659 | 0,617 | 0,659 | 0,626 |
| Consensus | 0,603 | 0,518 | 0,603 | 0,289 |
| Social + Consensus | 0,673 | 0,586 | 0,673 | 0,332 |
| Quantum Consensus | 0,611 | 0,577 | 0,615 | 0,584 |
| Social + Quantum Consensus | 0,675 | 0,624 | 0,691 | 0,637 |
| Quantum Social | 0,833 | 0,755 | 0,840 | 0,774 |
| Quantum Social + Consensus | 0,853 | 0,715 | 0,853 | 0,406 |
| Quantum Social + Quantum Consensus | **0,852** | **0,765** | **0,864** | **0,788** |

**Table. I.** Maximum and average STR for each operation mode in the first round (basic permafrost use case) and second round of tests (by following the generalization approach).

If we analyze the average STR results of these simulations (Table I), we can spot a big difference in the improvement of quantum-assisted mechanisms. In the first round of tests (Antarctica's use case), "Quantum Consensus", "Quantum Social", and "Quantum Social + Quantum Consensus" improved their analogous classical modes by 1,3%, 26%, and 39%, respectively. In contrast, we observe that in this second round (the generalization of IoT telemetry), the "Quantum Consensus" mode improves the "Consensus" mode by 2%, the "Quantum Social" mode improves the "Social" mode by 27%, and the "Quantum Social + Quantum Consensus" mode improves the "Social + Consensus mode" by 28%. These results remark that the use of the quantum-assisted mechanisms gains more relevance as the number of members in the redundancy increases. Thus, the quantum-assisted mechanisms enable broader scenarios with more sensors and measuring spots deployed.

Beyond the improvements introduced in [3] about the fast quantum consensus, and more specifically in the main contribution of this article regarding the quantum social trustworthiness control plane: The non-additivity property of quantum channels is a fundamental property of quantum mechanics that distinguishes it from classical physics. In classical physics, the capacity of a communication channel is determined by its physical properties, such as bandwidth and noise level. These properties are additive, which means that the combined capacity of two channels is the sum of their individual capacities. In contrast, quantum channels exhibit non-additivity, which means that the combined capacity of two quantum channels can be greater than the sum of their individual capacities. This property arises from the non-classical correlations between quantum systems, such as entanglement. The non-additivity property of quantum channels can be modeled as stated in the simulation setup chapter. It also highlights the non-classical properties of quantum mechanics and the potential for quantum-assisted technologies to achieve higher performance than classical technologies.

### ACKNOWLEDGMENT

This research was funded by the European Social Fund (ESF) 2022 FI_B2 00026. This work also received funding from the Spanish Ministry of Science, Innovation and University, the Investigation State Agency, and the European Regional Development Fund (ERDF) under the grant number RTI2018-097066-B-I00 (MCIU/AEI/FEDER, UE) for the project "NVIS Sensor Network for The South Shetland Islands Archipelago" (SHETLAND-NET).

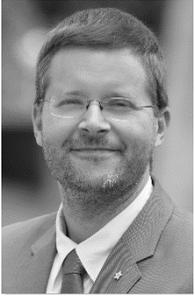

**Agustín Zaballos** was born in Madrid, Spain, in 1976. He received the M.S. and Ph.D. degrees in electronics engineering from La Salle URL (Universitat Ramon Llull), in 2000 and 2012, respectively. He also holds an International MBA from La Salle URL since 2014. He is a Full Professor in the Computer Science Department since 2017 and is the head of the Research Group in Internet Technologies (GRITS). He is currently the Commissioner of BitLaSalle (Barcelona Institute of Technology LaSalle), in charge of the strategy regarding multidisciplinary research projects in La Salle Campus Barcelona.

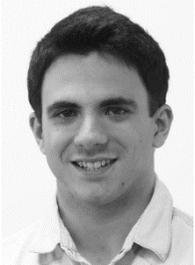

**Adrià Mallorquí** was born in Calella, Barcelona, Spain in 1995. He received the M.S. and Pd.D. degrees in network and telecommunication engineering from La Salle URL (Universitat Ramon Llull), in 2019 and 2023, respectively. From 2014 to 2017, he was a Research Intern in the Research Group on Internet Technologies (GRITS). In 2017 he became a Research Assistant at the same group and an Assistant Professor in the Engineering Department.

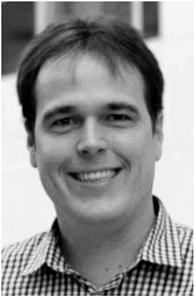

**Joan Navarro** was born in Barcelona, Spain, in 1985. He received his Ph.D. in Information and Communication Technologies from La Salle URL (Universitat Ramon Llull), in 2015, his MSc. degree in Telecommunications Engineering in 2008, and his BSc in Network Engineering in 2006. He is currently an Associate Professor at the Computer Engineering Department of La Salle (URL) and a member of the Group of Research in Internet Technologies (GRITS) of the same University. His research is focused on Cloud Computing, Big Data and Distributed Systems, specifically on the areas of Concurrency Control in Large-Scale Distributed Systems and Replication Policies in Cloud-based Databases.